\documentstyle[prl,aps,psfig,floats]{revtex}
\begin{document}
\draft
\twocolumn[
\hsize\textwidth\columnwidth\hsize\csname@twocolumnfalse\endcsname
\title{Data Analysis of Continuous Gravitational Wave Signal: Fourier Transform}
\author{D.C. Srivastava$^{1,2}$ and S.K. Sahay$^1$}
\address{$^1$Department of Physics, DDU. Gorakhpur University, Gorakhpur-273009,
U.P, India  \\ $^2$ Visiting Associate, Inter-University Centre for Astronomy and Astrophysics,
Post Bag 4, Ganeshkhind, Pune 411007, India}
%\date{\today}
\maketitle
\begin{abstract}
We present the Fourier Transform of continuous gravitational wave for
arbitrary location of detector and source and for any duration of observation
time in which both rotational motion of earth about its spin axis and orbital
motion around sun has been taken into account. We also give the method to
account the spin down of continuous gravitational wave.

\end{abstract}

% for PACS codes, see http://publish.aps.org/PACS/pacs99.html 
 %Gravitational wave detectors and experiments
\bigskip  
]\renewcommand{\thefootnote}{\arabic{footnote}} \setcounter{footnote}{0}
\narrowtext  
\par The detection of gravitational waves (GW) from their possible sources, at
 the present stage, has to largely depend upon the careful study of the 
expected characteristic of potential sources and the waveforms. In this 
context, the continuous sources, in particular pulsars, are now receiving much 
attention [1,2]. These may be the one of the sources to look for the signals in 
data 
streams of the long arm laser interferometer and are also attractive sources 
for currently operating bar detectors. The theoretical research clearly shows 
that the data analysis will involve processing of very long time series and 
exploration of a very large parameter space. It turns out that optimal 
processing of month long data streams is computationally prohibitive [1] and 
methods to reduce the computation burden are being worked out [4,5].
\par At the root cause of the excessive computational requirement lies the 
fourier transform (FT) analysis of the data. In order to detect the signal 
from the dominant noise one has to analyse the long time observation data ranging 
from months to year. The output of detector will be Doppler modulated; 
both frequency and amplitude due to the 
rotations of earth. The maximum Doppler shift for one kHz signal is
$\simeq  21.42 \times 10^{-2}$ [8].
 The output depends in a complex manner on the location of 
the source and the detector. 
Consequently, the FT is obtained using numerical techniques; one commonly in 
use is fast fourier transform (FFT). The evalutaion of FT analytically is an
important step in the data analysis and in contrast to FFT has following
advantages:
\begin{itemize}
\item The computational time is extremely less and particularly in the case
of long data train gains significance.
\item We can achieve the resolution of the transform as we need, hence dominant
peaks of doppler modulated signals can be accumulated, and
\item To make templates for matched filtering in colored noise
\end{itemize}
\par In the literature very little attention
has been paid to analytical analysis for fourier transform [6-8]. In this 
letter we address to this problem in its complete generality.   

The response $R(t)$ of the ground based laser interferometric detector at 
time $t$ is a linear combination of 
the two polarizations $h_+$ and $h_\times$ of the signal given via
\begin{equation}R(t) = F_+ ( t ) h_+ ( t ) +  F_\times ( t ) h_\times ( t )
\end{equation}
\noindent where $ F_+ $, $ F_\times $ are the beam pattern functions and 
represent the amplitude modulation (AM) part. In Solar System Barycentre (SSB) frame the beam 
pattern as function of the direction of the incoming wave $( \theta , \phi , 
\psi ) $, the orientation of the detector angles $( \alpha , 
\beta , \gamma )$ and $\epsilon $, the angle between the equator and the 
ecliptic may be expressed as [6]
 \begin{eqnarray}
F_+ & = &{1\over 2}\left[ 2 ( L^2 - M^2 ) Z U + ( N^2 - P^2 ) A \: - \right.\nonumber\\
&&\left. ( Q^2  - R^2) B \right] + ( L N - M P ) C + ( L Q + M R ) D  \nonumber\\
&& + ( N Q + P R ) E\, ,\\
F_\times & = & 2 L M Z U + N P A - {1\over 2} B \sin^2\theta\sin 2\phi\: +\nonumber \\
&& ( L P + M N ) C + ( M Q - L R ) D \nonumber \\
&& + ( P Q - N R ) E  
\end{eqnarray} 
\noindent where 
\begin{eqnarray*}
A &= &2 X Y\cos^2\epsilon - \sin^2\epsilon \sin^2\alpha \sin 2\gamma + \\
&& \sin 2\epsilon ( X \sin \alpha \sin \gamma - Y \sin \alpha \cos \gamma ) \, ,\\
B& = & 2 X Y\sin^2\epsilon - \cos^2\epsilon \sin^2\alpha \sin 2\gamma - \\
&&\sin 2\epsilon ( X \sin \alpha \sin \gamma - Y \sin \alpha \cos \gamma )  \, ,\\
C& =& \cos\epsilon ( Y V + X U ) + \sin\epsilon ( V \sin \alpha \sin
\gamma  - U\sin\alpha \cos \gamma )\, ,\\
D &= &- \sin\epsilon ( Y V + X U ) + \cos\epsilon ( V \sin \alpha 
\sin\gamma - U \sin\alpha \cos \gamma ]\, ,\\
E& = &- 2 X Y \cos\epsilon \sin\epsilon - \cos\epsilon\sin\epsilon
\sin^2\alpha\sin 2\gamma + \\
&& \cos 2\epsilon ( X \sin\alpha\sin\gamma - Y \sin \alpha \cos \gamma  )\, ,\\
L &= &\cos\psi\cos\phi - cos\theta\sin\phi\sin\psi\, ,\\
 M &=& \cos\psi\sin\phi + cos\theta\cos\phi\sin\psi\, , \\ 
N &= & - \sin\psi\cos\phi - cos\theta\sin\phi\cos\psi\, ,\\
 P& = &- \sin\psi\sin\phi + cos\theta\cos\phi\cos\psi \, ,\\
Q &= &\sin\theta\sin\phi\, ,\qquad  R\; =\; \sin\theta\cos\phi \, ,\\
U &= &- \cos \alpha \cos \beta \sin \gamma - \sin \beta \cos \gamma \, ,\\ 
V &= &\cos \alpha \cos \beta \cos \gamma - \sin \beta \sin \gamma\, , \\
X &= &\cos \alpha \sin \beta \cos \gamma + \cos \beta \sin \gamma \, ,\\
Y &=& - \cos \alpha \sin \beta \sin \gamma + \cos \beta \cos \gamma \, ,\\
\beta &=& \beta_o + w_r t 
\end{eqnarray*}

\noindent Here $\beta_o$ represent the initial azimuth of the detector. We 
have assumed that the arms of the laser interferometer are mutually 
orthogonal. However, if the angle between the arms of the detector are 
different than 
$90^o$ the expression for $F_+$ and $F_\times $ will simply get multiplied by 
sine of the angle between the arms of the detector respectively. Inclusion of 
this parameter does not pose  any difficulty as it simply affect the norm of 
the FT by the same factor.\\ 

After algebraic manipulation eq. (2) and (3) may be expressed as 

\begin{eqnarray}F_+ &= &F_{1_+}\cos 2\beta + F_{2_+}\sin 2\beta + F_{3_+}\cos \beta + \nonumber \\
&& F_{4_+}\sin \beta + F_{5_+}\\
F_\times &= &F_{1_\times}\cos 2\beta + F_{2_\times}\sin 2\beta + F_{3_\times}\cos \beta +  \nonumber \\
&&F_{4_\times}\sin \beta + F_{5_\times}\end{eqnarray}
\noindent $F_{i _+}$ and $F_{i_\times}$ $( i = 1,2,3,4,5 )$ are time independent and are given by
\begin{eqnarray*}
F_{1_+} &= & - 2 G \cos\alpha\cos 2\gamma + {H \sin 2\gamma\over 2}( \cos^2\alpha + 1 )  \, ,\\
F_{2_+} &= &H \cos\alpha\cos 2\gamma + G \sin 2 \gamma (\cos^2\alpha + 1 )\, ,\\
F_{3_+} &= &I\sin\alpha\cos 2\gamma + J \sin 2\alpha\sin 2\gamma\, ,\\
F_{4_+} &= &2 J \sin\alpha\cos 2\gamma - {I\over 2}\sin 2\alpha\sin 2\gamma\, ,\\
F_{5_+} &=& {3\sin^2\alpha\sin 2\gamma\over 2}[ H + L^2 - M^2 ]\, ,\\
G& =& {1\over 2}[ ( L Q + M R )\sin\epsilon - ( L N - M P )\cos\epsilon ]\, ,\\
H &=& {1\over 2}[ ( N^2 - P^2 ) \cos^2\epsilon - ( L^2 - M^2 ) + \\
&& ( Q^2 - R^2 )\sin^2\epsilon  - ( N Q + P R )\sin 2\epsilon ]\, ,\\
I& = &{1\over 2}[ (Q^2 - R^2 ) \sin 2\epsilon - ( N^2 - P^2 )\sin 2\epsilon -\\
&& 2 ( N Q + P R )\cos 2\epsilon ]\, ,\\ 
J& =& {1\over 2}[ ( L N - M P )\sin\epsilon + ( L Q + M R )\cos\epsilon ]
\end{eqnarray*}
\noindent and $ F_{i_\times} $ is related to $ F_{i_+} $ via
\begin{eqnarray} 
F_{i_\times}( \theta , \phi , \psi , \alpha , \beta , \gamma , \epsilon )& = & F_{i_+}( \theta , \phi - {\pi\over 4}, \psi , \alpha , \beta , \gamma , \epsilon )\\
&&  \qquad i\; = \;1,2,3,4,5 \nonumber 
\end{eqnarray}
The two polarisation states of the signal may be taken as
\begin{eqnarray*}
h_+(t)&=& h_{o_+}\cos [\Phi (t)] \, ,\qquad h_\times (t)\;=\; h_{o_\times}\sin [\Phi (t)] 
\end{eqnarray*}
where  $h_{o_+}$, $h_{o_\times}$ are the time independent amplitude of 
$h_+(t)$, and  $ h_\times (t)$. Here $\Phi(t)$ is phase of the GW signal and 
may be expressed in SSB frame as [7]

\begin{eqnarray}
\Phi (t) & = &  2\pi f_o \left[ t + {R_{se}\over c} \sin
\theta\cos\phi' + \right.\nonumber \\
&& {R_e\over c}\sin\alpha \{\sin\theta (\sin\beta\cos\epsilon\sin\phi + \cos\phi\cos\beta) + \nonumber \\
&&\sin\beta\sin\epsilon\cos\theta\} - {R_{se}\over c}
\sin\theta\cos\phi_o + \nonumber \\
&& {R_e\over c}\sin\alpha\{\sin\theta (\sin\beta_o\cos\epsilon\sin\phi 
 +  \cos\phi\cos\beta_o) +\nonumber\ \\
&&\left. \sin\beta_o\sin\epsilon\cos\theta\} \right]
\end{eqnarray}
\noindent where, $\phi' = w_{orb} t - \phi $, $R_{se}$ is the distance from 
the centre of the SSB frame to the centre of the earth, $R_e$ is the radius 
of the earth, $w_{orb}$ is the angular 
velocity of the earth around the sun, $w_r$ is the angular  
velocity of the earth about its spin axis and $c$ is the velocity of light. 
Here $f_o$ is the frequency of the continuous signal. \\ \\
\noindent The complete response of the detector is given via
\begin{eqnarray} \tilde {R}(f) &= &\tilde{R}_+(f) + \tilde{R}_\times (f) \, ,\\
\tilde{R}_+ (f) &= &\int h_+(t) F_+(t) e^{-i 2 \pi f t} dt \, ,\\
 \tilde{R}_\times (f)& = &\int h_\times (t) F_\times (t) e^{-i 2 \pi f t} dt\end{eqnarray}

\noindent Due to the symmetries involved in $F_+$ and $F_\times$ as given by
[eqn. (6)] it is sufficient
to evaluate either of the FT{\footnotesize s} (9) and (10) and the other may be obtained in a 
simple manner. We will, at present, consider $ \tilde{R}_+ (f)$ and get 

 \begin{eqnarray}
\tilde{R}_+(f) & = & h_{o_+}\left[e^{-i 2 \beta_o}( F_{1_+} + i F_{2_+} ) 
\tilde{h}_+ ( f + 2 f_r)/2 + \right. \nonumber \\
&& e^{i 2 \beta_o}( F_{1_+} - i F_{2_+} )\tilde{h}_+ ( f - 2 f_r)/2 + \nonumber \\
&& e^{-i \beta_o}( F_{3_+} + i F_{4_+} )\tilde{h}_+ ( f +  f_r)/2 + \nonumber \\
&& \left. e^{i \beta_o} ( F_{3_+} - i F_{4_+}) \tilde{h}_+ ( f -  f_r)/2 + F_{5_+}\tilde{h}(f)\right]
\end{eqnarray}
where
\begin{eqnarray}
\tilde{h}_+ (f) &= &\int h_{o_+}\cos[\Phi (t)]
\end{eqnarray}
and represent the FT arising due to frequency modulation. This means that the 
AM results in the four side bands at frequencies $f \pm f_r$ , $f \pm 2f_r $ 
around the central frequency $f$. For the moment, we keep the data interval of 
the observation time completely arbitrary. We evaluate the FT

\begin{equation}
\tilde{h}(f) \;= \;\tilde{h}_+ (f) \;= \;\int_{n\bigtriangleup t}^{( n + 1 )\bigtriangleup t} h_{o_+}\cos [\Phi (t)]
\end{equation}

\noindent where $n$ is an integer and $\bigtriangleup t$ is time interval. It 
is important to point out that it is not easy to evaluate the FT analytically 
and it is the stage where one has to take recourse to numerical methods. 
 The FT [eq. (13)] after lengthy but straight forward calculation is obtained
 as a double series 

\begin{eqnarray} 
\tilde {h}(f)& =& {h_{o_+} \nu \over 2 w_r} \sum_{k  =  - \infty}^{k =  
\infty} \sum_{m = - \infty}^{m =  \infty} e^{ i {\cal A}}{\cal B} ( {\cal Z},{\cal N} ) \left[ {\cal C } (\lambda , \zeta , \tau)\: - \right.\nonumber \\
&& \left. i {\cal D} (\lambda , \zeta , \tau )\right] 
\end{eqnarray}
where
\begin{eqnarray*}
\nu & = & ( f_0 - f )/f_r \, , \qquad ( f_r = w_r/2\pi ) \, ,\\
{\cal A}&  = &{(k + m)\pi\over 2} - {\cal R}\, , \\
{\cal B}( {\cal Z},{\cal N} ) & = & {J_k({\cal Z}) J_m({\cal N})\over {\nu^2 - (a k + m)^2}} \, , \\
{\cal C}(\lambda , \zeta , \tau ) &= &  \sin (\nu \tau )\cos ( a k \tau  + m \tau - k \lambda - m \zeta ) \\
&&- { a k + m \over \nu}\{\cos (\nu \tau ) \sin ( a k \tau + m \tau - k \lambda  - m \zeta ) \\
&& + \sin ( k \lambda + m \zeta )\}\, , \\ 
 {\cal D}(\lambda , \zeta , \tau ) & = & \cos (\nu \tau )\cos (  a k \tau +  m \tau - k \lambda - m \zeta ) \\
&& + {a k + m \over \nu}\sin (\nu \tau ) \sin ( a k \tau + m \tau - k \lambda - m \zeta )\\
&&  - \cos ( k \lambda + m \zeta) \, , \\
{\cal R} &=& {\cal Z}\cos\phi + {\cal Q} + 2\pi f_o n\bigtriangleup t\, ,\\
{\cal Z}& =& (2\pi /c) f_o R_{se}\sin\theta \, , \quad {\cal N}\; =\;  \sqrt{ {\cal P}^2 + {\cal Q}^2 } \, ,\\
\lambda & = & \phi -  a n \tau \, , \qquad \zeta \; = \; \delta -  n\tau \, ,\\
\delta & = & tan^{-1}\frac{{\cal P}}{{\cal Q}}\, , \qquad\tau = w_r\bigtriangleup t \, , \quad a = 1/365.25 \, ,\\
{\cal P} &= & (2\pi /c) f_o R_e \sin\alpha \left[\cos\beta _o(\sin\theta
\cos\epsilon \sin\phi\: + \right. \\
&&\left. \cos\theta \sin\epsilon ) - \sin\beta _o \sin\theta\cos\phi )\right] \, ,\\
{\cal Q} &=& (2\pi /c) f_o R_e\sin\alpha \left[\sin\beta _o (\sin\theta 
\cos\epsilon \sin\phi\: + \right.\,  \\ 
&& \left.\cos\theta \sin\epsilon ) + \cos\beta _o \sin\theta\cos\phi \right]\, ,
\end{eqnarray*}
\noindent $J_k({\cal Z}) , \; J_m({\cal N})$ represent the bessel function of first 
kind of integer order $k$ and $m$ respectively.\\

\noindent A case of particular interest arises where one needs to evaluate 
the FT of the data for $T_{obs}$ time i.e. the limits in (14)  are taken as 
$0$ to $T_{obs}$. The result in this case is obtained by taking\\ 

$n = 0$ $\, , \quad$ $\bigtriangleup t = T_{obs}$ $\quad \Longrightarrow  
\quad$ $\tau = w_rT_{obs}$\\ \\
Thus we obtain

\begin{eqnarray} 
\tilde{h}(f)& = & {h_{o_+} \nu \over 2 w_r} \sum_{k  =  - \infty}^{k =  
\infty} \sum_{m = - \infty}^{m =  \infty} e^{ i {\cal A}}{\cal B} ( {\cal Z},{\cal N} ) \left[ {\cal C } (\lambda , \zeta , w_rT_{obs} )\: - \right. \nonumber \\
&&\left.  i {\cal D} (\lambda , \zeta , w_rT_{obs} )\right]
\end{eqnarray}

\noindent Suppose one is interested in evaluating the FT for a day or year 
observation data then he/she can obtain the FT by putting $n = 0$ in eq. (14),
 and \\ \\
$T_{obs}$ = one year $\quad \Longrightarrow  \quad$  $\tau = 2\pi /a$ \\ \\
$T_{obs}$ = one day $\quad \Longrightarrow  \quad$  $\tau = 2\pi$  \\ 

\par The result obtained [eq. (14)] is useful in getting the FT of the complete
response of a detector of GW being emitted by a pulsar whose frequency is
either slowing down or picking up. To account for this aspect, we consider
 the evaluation in 
different window of time by splitting the interval ($0$, $T_{obs}$) in N equal
 parts, each of interval $\bigtriangleup t$ $(T_{obs} = N\bigtriangleup t$) 
such that the signal over a window may be treated as monochromatic. The
strategy is to evaluate the FT over the window and finally to add the result.
This process has also been suggested by Brady \& Creighton [6] and Schutz [4] in numerical computing and
called by them as stacking.
The transform [eq. (14)] can be computed easily for data of any observation 
time. Consequently, the obtained transform may be one of the efficient tool 
for the detection of continuous gravitational wave. To obtain the 
result one has to evaluate $\tilde{R}(f)$ in each window by accounting for 
changing 
value of $f_o$.\\ 
\par The authors are thankful to Prof. S. Dhurandhar, IUCAA, Pune for stimulating 
discussion. This work is supported by research grant number SP/S2/0-15/93 by 
DST, New Delhi. S.K. Sahay is also thankful to IUCAA for providing hospitality during 
the course of stay at IUCAA where major part of work was carried out. He is also
thankful to Prof. A. Kembhavi, IUCAA for encouragement.

\end{document}